\documentclass[preprint,12pt]{elsarticle}

\usepackage{epsfig}
\usepackage{amssymb}
\usepackage{amsmath}
\usepackage{amsthm}
\usepackage{tabularx}
\usepackage{booktabs}
\usepackage{graphicx}
\usepackage{float}
\usepackage{multirow} 
\usepackage{amsmath}
\usepackage{placeins} 
\usepackage{makecell} 
\usepackage{multirow}
\usepackage{xcolor}
\usepackage{tabularx} 
\usepackage{array} 
\usepackage{makecell}
\usepackage{verbatim}
\usepackage{lineno}
\usepackage{url}
\usepackage{xcolor} 
\begin{document}
\begin{frontmatter}

\title{Data-Driven Greenhouse Climate Regulation in Lettuce Cultivation Using BiLSTM and GRU Predictive Control}

\author[1]{Soumo Emmanuel Arnaud$^{,*}$}
\author[3]{Marcello Calisti}
\author[1]{Athanasios Polydoros}

\affiliation[1]{organization={University of Lincoln, School of Engineering, Physical Science and Lincoln Centre for Autonomous Systems},
            addressline={}, 
            city={Lincoln},
            postcode={LN6 7TS}, 
            state={},
            country={United Kingdom}}

\affiliation[3]{organization={Sant'Anna School of Advanced Studies, The BioRobotics Institute},
            addressline={Viale Rinaldo Piaggio 34}, 
            city={Pontedera},
            postcode={}, 
            state={Pisa},
            country={Italy}}

\begin{abstract}

Efficient greenhouse management is essential for sustainable food production, but the high energy demand for climate regulation poses significant economic and environmental challenges. While traditional process-based greenhouse models exist, they are often too complex or imprecise for reliable control. To address this, our study introduces a novel data-driven predictive control framework using Long Short-Term Memory (LSTM) and Gated Recurrent Unit (GRU) neural networks within a Model Predictive Control (MPC) architecture. Training data were generated from a validated dynamic model simulating lettuce cultivation under various environmental conditions. The LSTM and GRU networks were trained to predict future greenhouse states---including temperature, humidity, CO\textsubscript{2} concentration, and crop dry matter---with robustness confirmed via $10$-fold cross-validation. These networks were embedded into an online MPC controller to optimize heating, ventilation, and CO\textsubscript{2} injection, aiming to minimize energy consumption and maximize crop yield while respecting biological constraints. Results showed that both the LSTM- and GRU-based controllers significantly outperformed a conventional MPC baseline. For example, humidity violations dropped from 54.77\% (MPC) to 15.45\% (GRU) and 17.71\% (LSTM), while day--night temperature deviations were kept below $2^\circ\text{C}$. The GRU controller further achieved up to 40\% lower computation time than its LSTM counterpart, confirming its real-time feasibility. Overall, the proposed GRU-driven predictive control approach offers a robust and computationally efficient solution for intelligent greenhouse climate automation under practical operational constraints.

\end{abstract}


\end{frontmatter}
\section{INTRODUCTION}
Global food security faces significant challenges. The number of undernourished people has increased steadily since 2014, reaching an estimated 720 to 811 million in 2020 \cite{food2022food}. This rise occurs as the world population, currently around 7.7 billion, is projected to reach 9.7 billion by 2050, driving a rapid increase in the demand for fresh and nutritious food \cite{nations2019growing}. Consequently, food production must increase by approximately 70\% between 2005 and 2050 to meet future needs \cite{fao2018future}. At the same time, critical resources such as freshwater \cite{marcelis2019achieving}, fossil fuels, and arable land are becoming increasingly scarce, while the impacts of climate change continue to intensify. As a result, producing high-quality fresh food in a resource-efficient manner is more important than ever. 
Greenhouse cultivation has emerged as a powerful approach to agricultural intensification. Studies indicate that it can achieve substantial yield increases, typically ranging from 10\% to 20\%, while simultaneously reducing critical resource inputs by 25\% to 35\% \cite{stanghellini2013horticultural,de2012overview}. These considerable benefits have contributed to the accelerated global growth of protected horticulture. Nevertheless, the widespread adoption of greenhouse technologies is limited by inherent challenges, notably their high energy demands and the significant initial capital investments required \cite{graamans2018plant}.
\\
Greenhouses are essential for crop cultivation, as they provide a controlled environment that shields plants from unpredictable outdoor weather. In today’s world where energy costs are rising and labor and resources are becoming increasingly scarce, efficient and automated greenhouse management is more important than ever. This approach helps maximize crop production while minimizing resource use. However, outdoor weather constitutes a significant disturbance, making it challenging to control greenhouse climate, develop accurate models, and optimize operations for optimal performance \cite{van2010optimal}.
A greenhouse is a complex system consisting of two primary components: the cultivated plants and the surrounding controlled environment. This internal environment is influenced by a combination of external weather conditions, the functioning of greenhouse equipment, and the physiological activities of the plants themselves. To understand, predict and optimize the behavior of such systems, researchers commonly rely on mathematical modeling. These models serve a dual purpose: they offer information on the underlying physical and biological processes and support the development of control strategies to improve system performance \cite{lopez2018development,katzin2022process}.
\\
Traditional greenhouse modeling approaches are typically deterministic and based on first-principles, drawing from established scientific understanding of plant physiology, thermodynamics, and fluid dynamics \cite{norton2007applications,choab2019review}. Although these models can provide detailed and mechanistic insights, they are often complex, featuring nonlinear dynamics and a large number of parameters that have to be manually tuned. This complexity reduces their adaptability to different greenhouse setups or operational conditions. Furthermore, first-principles models are generally limited in their ability to handle uncertainty, as estimating the variability in model parameters requires extensive datasets and computationally demanding statistical methods. Although these models use data for calibration, they are not data-driven like machine learning models. As a result, they lack the flexibility to adapt to new conditions without manual adjustments. Building upon these modeling approaches, automated control systems have been increasingly adopted in greenhouse operations to maintain optimal environmental conditions \cite{maraveas2023agricultural}. Among advanced techniques, the MPC-controller stands out for its superior performance and energy efficiency compared to traditional methods \cite{afram2014theory}. The prediction model forms the cornerstone of any MPC strategy; its accuracy directly influences the quality of future state predictions and, consequently, the effectiveness of control actions. MPC typically employs analytical models—often in state-space form—to forecast system dynamics and determine the optimal control inputs over a specified prediction horizon \cite{bersani2020model}. However, in practical applications, the development of first-principles models is frequently hindered by challenges such as incomplete system knowledge, complex parameter identification, and computational burdens that are incompatible with real-time operation. This has led to the emergence of data-driven black-box models as a viable alternative within the MPC framework.  These models, trained exclusively on system input and output data, offer enhanced adaptability and computational efficiency.
However, existing machine learning applications in greenhouse climate control have shown considerable promise but still face several fundamental challenges. Models often struggle to generalize under unseen weather conditions, exhibit sensitivity to sensor noise, and require computational resources that hinder real-time deployment \cite{hosseini2025machine}. At the core of these issues lies a critical trade-off between maintaining expressive capacity for capturing long-term temporal dependencies and ensuring computational efficiency suitable for continuous control tasks. Although Transformer-based architectures are a dominant choice in many time-series applications, their substantial computational overhead and mixed performance evidence—where recurrent models such as GRUs and LSTMs have demonstrated superior accuracy \cite{feng2024were,lian2024comparative} in certain contexts make them less suitable for the present application.
\\
Recurrent neural networks (RNNs)—particularly LSTM and GRU architectures—address several of these challenges through their ability to model nonlinear relationships and retain long-range temporal dependencies within sequential data. Unlike static regression models, LSTM and GRU networks can dynamically adapt to evolving system conditions, providing improved forecasting accuracy and robustness to environmental fluctuations \cite{ahmed2023deep}. Moreover, their compact architecture enables real-time deployment when properly optimized, making them attractive candidates for integration within MPC frameworks.
\\
Despite these advantages, their systematic evaluation for real-time greenhouse climate control remains limited. Few studies have rigorously assessed the predictive accuracy, robustness, and computational feasibility of LSTM- and GRU-based models within operational control settings. This research aims to fill this gap by integrating and comparing LSTM and GRU black-box models within an MPC framework for greenhouse control. The study investigates their performance in terms of predictive capability, resilience to environmental disturbances, and suitability for real-time implementation under realistic operational constraints and varying prediction horizons.
\\
The remainder of this paper is organized as follows. Section II reviews related work on data-driven control approaches for greenhouse systems. Section III presents the experimental setup and dataset used in this study. Section IV introduces the architecture and performance evaluation of the proposed LSTM and GRU predictive models. Section V describes the integration of these models within a MPC framework. Section VI reports and discusses the simulation results, highlighting the control performance. Finally, Section VII concludes the paper and outlines directions for future research.

\section{Related Work}

Deep learning has become an increasingly prominent tool in time series forecasting due to its ability to model complex nonlinear dynamics and automatically extract hierarchical representations from raw data—capabilities that conventional statistical and mechanistic models often lack \cite{li2021towards,torres2021deep}. Its proficiency in handling large-scale, high-dimensional, and multi-modal datasets has positioned deep neural networks as powerful alternatives for environmental and agricultural modeling tasks \cite{liu2024state}. In the context of greenhouse systems, recurrent neural architectures such as LSTM and GRU networks have been widely investigated for their capacity to capture long-term temporal dependencies and nonlinear variable interactions. For example, \cite{huang2024edible} proposed a hybrid architecture combining attention mechanisms, convolutional layers, and LSTMs to enhance the prediction of greenhouse temperature, humidity, and $CO_2$ concentrations, demonstrating significant improvements in accuracy. Likewise, \cite{he2022gated} showed that GRU networks can deliver high-fidelity temperature forecasts even when trained with limited input features, outperforming conventional data-driven methods.
\\
Parallel advancements in artificial intelligence have catalyzed the integration of deep learning models within control frameworks, particularly MPC, to enhance greenhouse climate regulation. Data-driven MPC approaches have demonstrated their potential to improve energy efficiency and maintain precise environmental control under uncertainty \cite{mahmood2023data}. The stability of recurrent models in such control settings has been examined in studies such as \cite{terzi2021learning}, which established sufficient conditions for Input-to-State Stability (ISS) in LSTM networks and leveraged these properties to design asymptotically stable observer-based MPC frameworks. Similarly, \cite{bonassi2021nonlinear} trained stable GRU networks for plant identification within nonlinear MPC, achieving offset-free reference tracking and stability guarantees validated on benchmark processes.
\\
Reducing the computational overhead associated with nonlinear MPC has also been a focus of research. Neural network-based surrogates have been developed to approximate system behaviors and streamline optimization processes. \cite{masti2020learning} introduced an approach using artificial neural networks to parameterize output predictions, facilitating efficient Linear Time-Varying MPC formulations for nonlinear systems. Building on this direction, \cite{doncevic2020deterministic} employed recurrent networks within MPC to lower computational requirements and demonstrated real-time capability through parallel computing architectures. Further innovations combine LSTM and GRU predictors with trajectory linearization techniques \cite{zarzycki2022advanced} or hybrid physics-informed strategies \cite{ye2023real}, achieving improved control performance and predictive precision across engineering applications, including motor drive regulation \cite{hammoud2022learning} and vehicle energy optimization \cite{baby2022data}.
\\
Although these studies collectively underscore the efficacy of deep recurrent models in both prediction and control, several limitations persist. Most existing works either isolate forecasting performance or evaluate control strategies under idealized or restricted scenarios. Comprehensive investigations that assess the real-time applicability, computational feasibility, and robustness of LSTM- and GRU-based models for greenhouse climate control remain limited. Consequently, there is a distinct need for systematic evaluation frameworks that bridge prediction accuracy and closed-loop control performance under practical operational conditions. The present study addresses this gap by comparatively analyzing deep recurrent architectures against traditional modeling approaches within an integrated MPC framework for greenhouse environments, thereby extending current knowledge on data-driven strategies for sustainable climate management.
\section{System Description and Data Acquisition}
\begin{figure*}[ht]
    \centering
    \includegraphics[width=\textwidth]{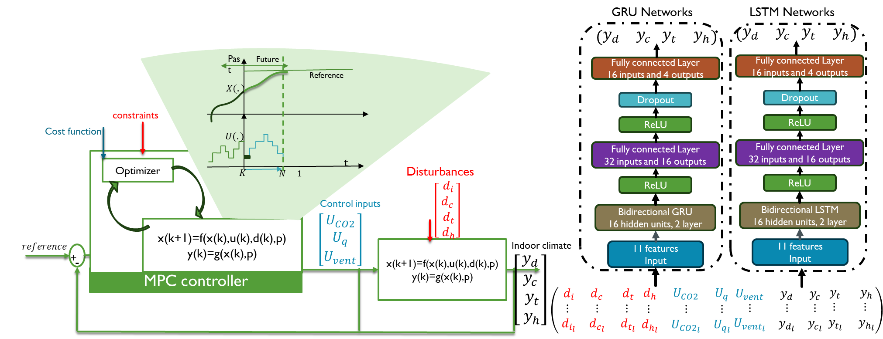}
    \caption{An overview of the MPC framework combined with LSTM and GRU neural networks is presented. The MPC controller generates control commands based on predictions from the van Henten \cite{van1994greenhouse} greenhouse model, external disturbances (solar radiation $d_i$, ambient temperature $d_t$, external CO$_2$ concentration $d_c$, and ambient humidity $d_h$), and reference trajectories. The LSTM and GRU networks are then trained on the data generated by this process. These networks learn to predict future greenhouse states—lettuce dry matter ($y_d$), indoor CO$_2$ concentration ($y_c$), indoor temperature ($y_t$), and indoor humidity ($y_h$)—based on historical data, disturbances, and previous control inputs. The  modeling capabilities of the LSTM and GRU units to support the optimization process within the MPC loop is investigated.}
    \label{fig:mpc_lstm_gru_framework}
\end{figure*}
\subsection{Data Acquisition}

Training data for both the LSTM and the GRU models were generated through simulations using a MPC strategy~\cite{morcego2023reinforcement} applied to the van Henten greenhouse model described in Section~\ref{Greenhouse_model}. Real-world datasets that simultaneously capture outdoor disturbances, greenhouse control inputs, and internal climate responses are limited. To address this scarcity, we simulated control trajectories by applying the MPC controller to the greenhouse model under diverse environmental conditions and greenhouse constraints. This approach enabled the creation of realistic and varied training scenarios that closely replicate real-world greenhouse operations while maintaining experimental control.
The data of outdoor weather disturbances, that consist of solar radiation (W·m$^{-2}$), air temperature (°C), CO$_2$ concentration (kg·m$^{-3}$) and absolute humidity (kg·m$^{-3}$), were obtained from long-term records collected between 2001 and 2020 at Schiphol Airport in the Netherlands~\cite{vangreenlight}  and are available online\footnote{\url{https://github.com/BartvLaatum/GreenLightGym}}. An overview of the dataset used in this study is presented in Table \ref{tab:data_overview}

\begin{table}[h!]
\centering
\caption{Overview of the Data used in this study}
\label{tab:data_overview}
\resizebox{\textwidth}{!}{%
\begin{tabular}{ll}
\hline
\textbf{Item} & \textbf{Value} \\
\hline
Original Sampling Frequency & 5-minute interval (300 s) \\
Resampling Frequency & 15-minute interval (900 s) \\
Number of missing values & 0 \\
Number of samples in the dataset & 701200 \\
Number of samples per fold (10-fold experiment) & 70120 \\
Data preprocessing & Normalized to the unit range \\
Outdoor Weather Data Location & Schiphol Airport, Netherlands (2001--2020) \\
Outdoor Weather Data Features & Solar radiation, outdoor temperature, \\
& CO$_2$ concentration, absolute humidity \\
\hline
\end{tabular}
}
\end{table}

\begin{figure}
    \centering
    \includegraphics[width=\textwidth]{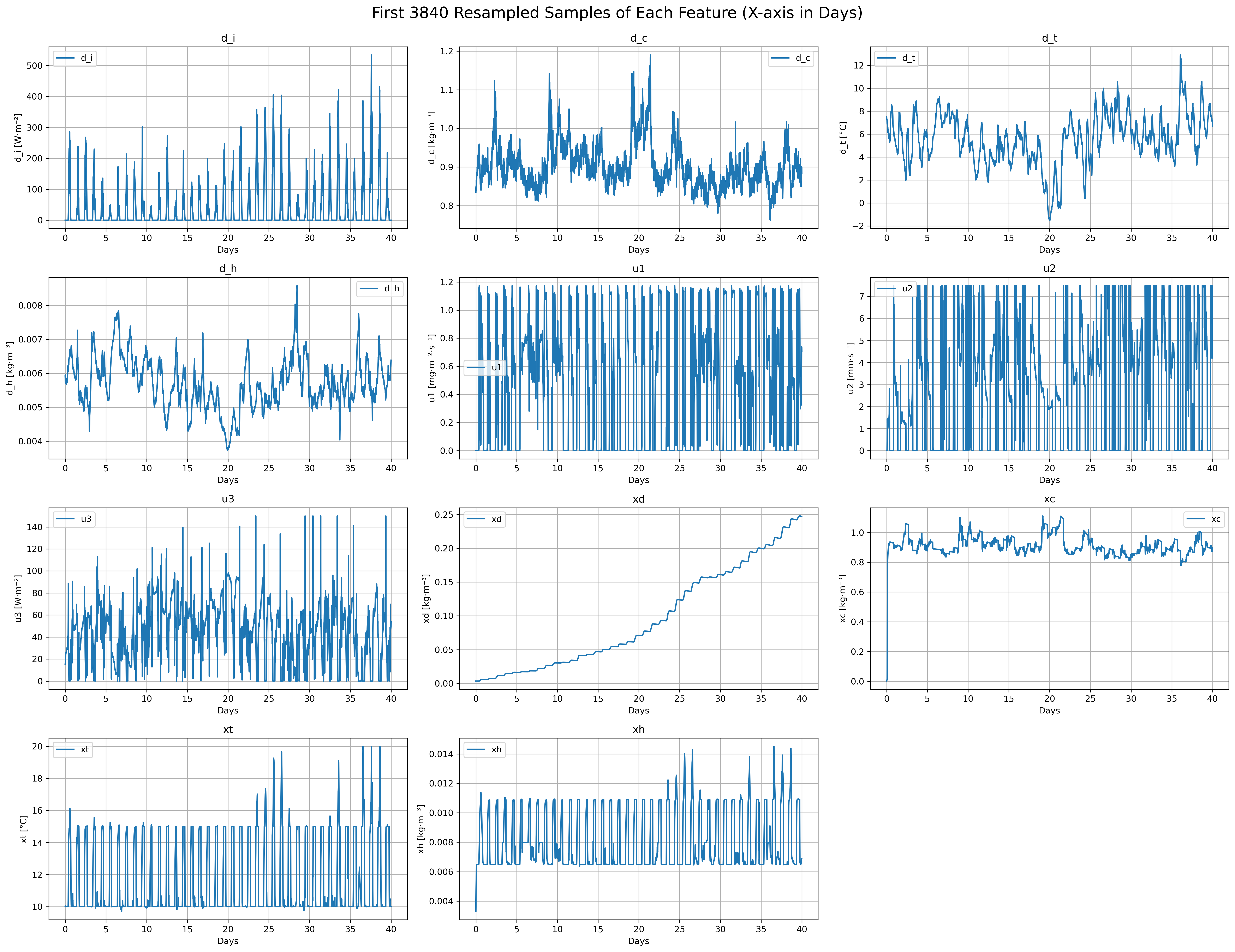}
    \caption{A sample of 3840 data points from the training dataset generated using the MPC strategy. The plot includes disturbances (solar radiation in W·m$^{-2}$, outdoor temperature in °C, outdoor CO$_2$ concentration, and outdoor humidity), control actions (CO$_2$ injection in mg·m$^{-2}$·s$^{-1}$, ventilation rate in mm·s$^{-1}$, and heating system output in W·m$^{-2}$), and greenhouse states (lettuce dry matter, indoor CO$_2$ concentration in kg·m$^{-3}$, indoor temperature in °C, and indoor humidity in kg·m$^{-3}$). Time is expressed in days. These data were used to train the LSTM and GRU models.}

    \label{fig:gru24_prediction}
\end{figure}

\section{Modeling and Control Framework}
\subsection{Predictive Model Based on LSTM and GRU Networks}
\label{model}
RNNs, such as LSTM and GRU, are well-suited for modeling sequential data due to their capability to capture temporal dependencies. In greenhouse climate modeling, where environmental dynamics unfold over time, these models are particularly useful. Among the RNN variants, LSTM networks stand out for their superior long-term memory through gating mechanisms and internal memory cells, allowing them to selectively retain and discard information.

At each time step \( t \), the LSTM network processes three inputs: the current input sequence \( x_t \), the previous hidden state \( h_{t-1} \), and the previous cell state \( C_{t-1} \). The update rules are as follows:

\begin{align}
i_t &= \sigma(W_i [h_{t-1}, x_t] + b_i) &\text{(input gate)} \\
\tilde{C}_t &= \tanh(W_c [h_{t-1}, x_t] + b_c) &\text{(candidate cell state)} \\
f_t &= \sigma(W_f [h_{t-1}, x_t] + b_f) &\text{(forget gate)} \\
C_t &= f_t \odot C_{t-1} + i_t \odot \tilde{C}_t &\text{(cell state update)} \\
o_t &= \sigma(W_o [h_{t-1}, x_t] + b_o) &\text{(output gate)} \\
h_t &= o_t \odot \tanh(C_t) &\text{(hidden state)} \\
y_t &= \phi(W_y h_t + b_y) &\text{(network output)}
\end{align}

where, \( \sigma(\cdot) \) and \( \tanh(\cdot) \) represent the activation functions of the sigmoid and hyperbolic tangent, respectively, \( \odot \) denotes the elementwise multiplication and \( \phi(\cdot) \) is the activation function of the output. Parameters \( W_* \) and \( b_* \) are learned using backpropagation through time (BPTT).

The GRU network simplifies the LSTM structure by combining the forget and input gates into a single update gate and merging the cell and hidden states. This results in fewer parameters and reduced training time:

\begin{align}
z_t &= \sigma(W_z [h_{t-1}, x_t] + b_z) &\text{(update gate)} \\
r_t &= \sigma(W_r [h_{t-1}, x_t] + b_r) &\text{(reset gate)} \\
\tilde{h}_t &= \tanh(W_h [r_t \odot h_{t-1}, x_t] + b_h) &\text{(candidate hidden state)} \\
h_t &= (1 - z_t) \odot h_{t-1} + z_t \odot \tilde{h}_t &\text{(hidden state)} \\
y_t &= \phi(W_y h_t + b_y) &\text{(network output)}
\end{align}
For both models, the input at each time step consists of external disturbances, past control inputs, and output measurements:

\[
\begin{aligned}
x_t = [&d(t), d(t-1), \dots, d(t - l_d); \\
       &u(t), u(t-1), \dots, u(t - l_u); \\
       &y(t), y(t-1), \dots, y(t - l_y)],
\end{aligned}
\]

where \( l_d \), \( l_u \), and \( l_y \) denote the number of historical time steps for disturbances, control inputs, and output variables, respectively. In this study, we systematically benchmark the effect of different historical window sizes ($l_d = l_u = l_y = \{6, 12, 18, 24\}$) on prediction accuracy. This input sequence is used to predict the immediate next output of the system, i.e., the system behavior 15 minutes ( sampling time: 900 seconds) ahead.
\subsubsection{Network Architecture}
The implemented LSTM and GRU networks share a similar architecture tailored to greenhouse climate prediction, each consisting of two bidirectional layers with a hidden size of 16 units per direction, effectively producing a 32-dimensional output per time step. The input to the networks has 11 features that represent external disturbances, control inputs, and output measurements. The output from the final time step of the recurrent layers is fed through two fully connected layers, where the first reduces the dimension from 32 to 16 with a ReLU activation and dropout, and the second maps to 4 predicted output features. This design enables effective sequence modeling of greenhouse climate dynamics for accurate prediction of the system's output.
\subsubsection{Training Objective and Experimental Setup}

The models were trained to minimize the mean squared error (MSE) over a single-step prediction horizon \( T_p = 1 \):

\[
\mathcal{L} = \left\| R(t) - \hat{Y}(t) \right\|_2^2,
\]

where \( R(t) = r(t+1) \) is the target value in the next time step, and \( \hat{Y}(t) = \hat{y}(t+1) \) denotes the prediction of the corresponding model.

We performed 10-fold cross-validation on the 20-year time series, using an 90/10 train-test (631080/70120 samples) split for each fold. In each fold, we trained both GRU and LSTM models by varying input window lengths (\(6, 12, 18, 24\)) and batch sizes (\(8, 16, 32\)). This approach resulted in \(240\) independent training experiments (\(10\) folds \(\times\) \(4\) window lengths \(\times\) \(3\) batch sizes \(\times\) \(2\) models). We evaluated model performance using the MSE and Root Mean Squared Error (RMSE), reporting the mean and standard deviation across all folds.
The generalizability test was designed to evaluate how well the best-performing GRU and LSTM configurations, trained during cross-validation, could adapt to unseen temporal segments of the 20-year series.

\subsubsection{Advantages} 
The primary benefits of using LSTM and GRU for predictive modeling in greenhouse systems are twofold: (1) their gating mechanisms enable the extraction of meaningful temporal features from sequential input data, which aligns well with the feedback nature of MPC, and (2) unlike classical predictive methods (e.g., AutoRegressive Integrated Moving Average (ARIMA), they do not rely on a fixed-order assumption, making them more robust and generalizable to various system dynamics.
\subsection{Online Model Predictive Control Framework}
In greenhouse climate management, key system states such as indoor air temperature, absolute humidity, and CO\(_2\) concentration are directly related to crop productivity. Maintaining these states within biologically safe and effective ranges is essential: extreme temperatures can halt plant growth, and excessive humidity fosters mold, reducing both yield and quality. As a result, constraints are applied to these states to safeguard plant health and optimize lettuce production. The primary control variables include heating, ventilation and CO\(_2\) enrichment, while external disturbances include solar radiation, ambient temperature, ambient humidity, and ambient CO\(_2\) levels.
Building on the LSTM and GRU models described in section \ref{model}, we implement a predictive control strategy using the respective trained LSTM and GRU networks.
The MPC controller operates in a receding-horizon manner. At every sampling instant \(k\), 
the controller receives measurements of the greenhouse states
\[
y(k) = [\, y_{d}(k),\, y_{\mathrm{CO}_2}(k),\, y_{T}(k),\, y_{H}(k) \,],
\]
and external disturbances
\[
d(k) = [\, d_{\mathrm{rad}}(k),\, d_{\mathrm{CO}_2}(k),\, d_{T}(k),\, d_{H}(k) \,].
\]

Using these inputs, the MPC solves an optimization problem over a prediction horizon 
\(N_{p}\). 
The BiLSTM or GRU model provides multi-step predictions of the greenhouse dynamics:
\[
\hat{y}(k+i+1) =
f_{\theta}\big( y(k+i),\, u(k+i),\, d(k+i) \big),
\qquad i = 0,\dots, N_{p}-1,
\]
where \(f_{\theta}\) denotes the trained recurrent model.

Thus, the neural networks replace the original van Henten model inside the MPC loop during 
prediction, enabling the controller to compute optimized control sequences using 
learned dynamics.
This predictive control follows a receding horizon approach in which the system state is measured at each time step and used to update the model's initial condition. The optimization problem is then solved on a finite prediction horizon \(N_p\), with the aim of minimizing a cost function that balances crop yield and energy consumption. Only the first control input from the optimized sequence is applied, after which the process is repeated at the next time step using the newly measured state.

The optimization problem seeks to determine the control inputs that minimize a weighted cost function:
\begin{equation}
V(u(k), y(k)) = -q_{y_d} \cdot y_1(k) + \sum_{j=1}^{3} q_{u_j} \cdot u_j(k),
\label{eq:cost1} 
\end{equation}
In this study, \( q_{y_d} = 1000 \) and \( q_{u_j} = \{10, 1, 1\} \) are tunable weights that were selected to balance the relative importance of tracking error and control effort in the cost function. This cost function captures the trade-off between maximizing lettuce yield and minimizing energy usage. The optimization is subject to the dynamic system behavior modeled by the trained LSTM or GRU networks, as well as constraints on control inputs, their rate of change, and time-varying output bounds. Notably, the indoor temperature constraints vary throughout the day to reflect crop-specific day--night requirements \cite{seginer1994optimal}. These constraints ensure that the controller operates within safe and biologically meaningful ranges while optimizing overall greenhouse performance.
The optimization problem is subject to the following constraints on control inputs and system outputs:

\begin{equation}
\begin{aligned}
&\mathbf{u}_{\min} = 
\begin{bmatrix}
0 \\
0 \\
0\\
\end{bmatrix},
\quad
\mathbf{u}_{\max} = 
\begin{bmatrix}
1.2 \\
7.5 \\
150 
\end{bmatrix},
\quad
\Delta \mathbf{u}_{\max} = \frac{1}{10} \cdot \mathbf{u}_{\max}, \\
&\mathbf{y}_{\min}(k) = 
\begin{bmatrix}
0 \\
0 \\
f_{y_t,\min}(k) \\
50 
\end{bmatrix},
\quad
\mathbf{y}_{\max}(k) = 
\begin{bmatrix}
\infty \\
1000 \\
f_{y_t,\max}(k) \\
85 
\end{bmatrix},
\end{aligned} 
\label{eq:cost2} 
\end{equation}

where the time-varying temperature bounds are given by:

\begin{equation}
f_{y_3,\min}(k) = 
\begin{cases}
10, & \text{if } d_i(k_0) < 10 \\
15, & \text{otherwise}
\end{cases},
\end{equation}
\\
\begin{equation}
\quad
f_{y_3,\max}(k) = 
\begin{cases}
15, & \text{if } d_i(k_0) < 10 \\
20, & \text{otherwise}
\end{cases}.
\label{eq:cost3} 
\end{equation}
Additionally, CO$_2$ injection is restricted to daytime hours. This is enforced by setting the third control input to zero outside the daytime:

\begin{equation}
u_{CO_2}(k) =
\begin{cases}
0, & \text{if } d_i(k_0) < 10, \\
u_{CO_2}(k), & \text{otherwise},
\end{cases}
\label{eq:cost4}
\end{equation}
These constraints ensure biologically appropriate greenhouse conditions, including variation in diurnal temperature and restricting CO$_2$ enrichment to periods when photosynthesis can actively benefit from it. The optimization problem described in equations~\ref{eq:cost1}--\ref{eq:cost4} is solved in \textsc{Python} using the open-source tools CasADi~\cite{andersson2019casadi}, the IPOPT solver~\cite{wachter2006implementation}, and the L4casadi framework~\cite{salzmann2024learning}.
\section{Experimental Results and Performance Evaluation}
In this experiment, we trained and evaluated both LSTM and GRU neural network models to predict greenhouse environmental conditions based on historical data and control inputs. The training dataset, derived from a real greenhouse model controlled by an MPC, was resampled at a 15-minute interval and normalized to the unit range. Both models were implemented in \texttt{PyTorch} with a bidirectional architecture consisting of two layers and a hidden size of 16 units. The input sequences included 11 features on sliding windows of various lengths (6, 12, 18, and 24), and the target outputs comprised four key greenhouse variables. The training process used Adam optimizer with a learning rate of $3 \times 10^{-5}$ and a StepLR scheduler, and training was conducted over 15 epochs.

The performance of the model was evaluated using MSE and RMSE metrics in a separate test set composed of MPC-generated trajectories as shown in Figure \ref{fig:gru24_prediction}. The evaluation involved inverse-scaling of the model output and comparing them to actual greenhouse measurements. 
According to Table~\ref{tab:gru_lstm_metrics}, the GRU model achieved its best performance with a window size of 18 and a batch size of 8, recording an MSE of 0.01700 and an RMSE of 0.13038. Similarly, the LSTM model performed optimally at the same configuration (window size of 18 and batch size of 8), yielding an MSE of 0.01797 and an RMSE of 0.13400. Across all tested configurations, the GRU consistently exhibited slightly lower MSE and RMSE values compared to the LSTM, indicating marginally superior predictive accuracy and generalization capability. This finding aligns with previous studies (e.g.,~\cite{yamak2019comparison}) that report GRUs often achieve comparable or better forecasting accuracy than LSTMs while offering greater computational efficiency due to their simpler gating mechanisms. In addition, as shown in Table \ref{tab:gru_lstm_metrics1}, a two-sample t-test was performed to compare the RMSE in the GRU and LSTM models. There was a significant difference in RMSE between GRU and LSTM across most configurations; t(df) < 0, p < 0.05 in 11 out of 12 comparisons, indicating that GRU yielded a lower mean RMSE than LSTM. The only non-significant difference was observed for BS = 8, WS = 24 (t(df) < 0, p = 0.056). Nonetheless, LSTMs may retain advantages in capturing longer-term dependencies within more complex temporal patterns, albeit at the cost of increased computational demand. Consequently, the selection between GRU and LSTM architectures should consider both performance requirements and computational resources for the intended application.

\begin{figure}[H]
    \centering
    \includegraphics[width=\textwidth]{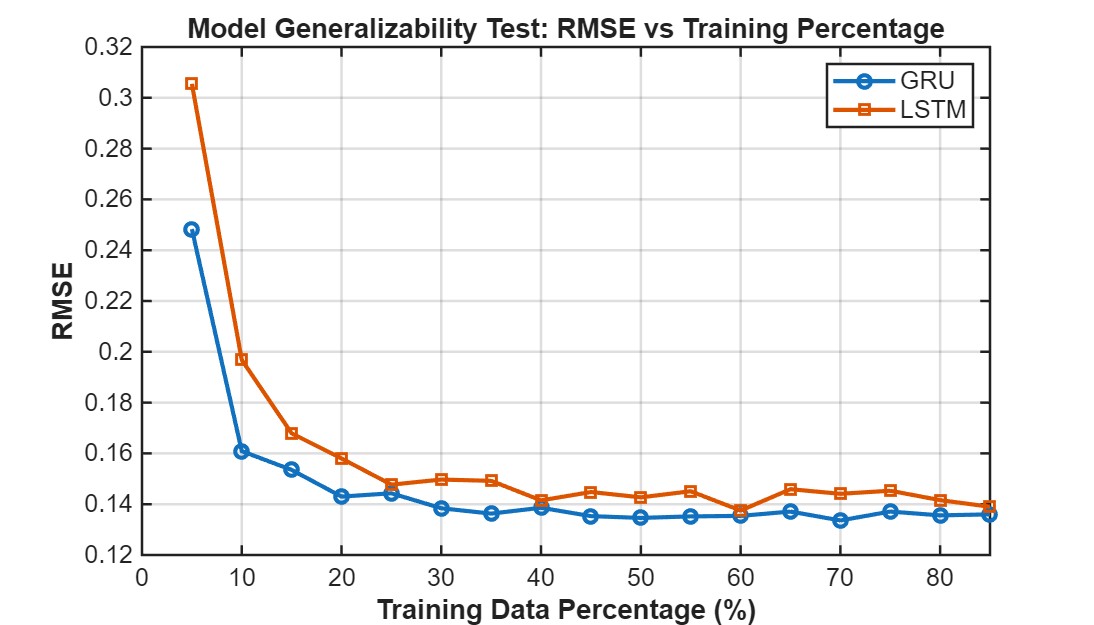}
    \caption{Variation of RMSE with training data percentage for GRU and LSTM models during the generalizability test.}
    \label{fig:3}
\end{figure}

These results highlight the importance of selecting an appropriate temporal context (input window) and batch size for optimal sequence modeling in data-driven greenhouse climate prediction. GRU appears to generalize slightly better than LSTM for this dataset, likely due to its simpler gating mechanism, which reduces the risk of overfitting when training on long temporal sequences. However, both models demonstrate strong generalizability across the 20-year dataset as shown in Figure~\ref{fig:3}, suggesting that either architecture can effectively capture greenhouse dynamics when properly tuned. Table~\ref{tab:gru_lstm_metrics} summarizes the results. The best-performing configurations for each model are highlighted in bold.
\begin{table}[H]
\centering
\caption{Mean $\pm$ Std of MSE and RMSE for GRU and LSTM models across different batch sizes (BS) and window sizes (WS). Bold values indicate the best performance (lowest error) for each model.}
\label{tab:gru_lstm_metrics}
\scriptsize
\resizebox{\textwidth}{!}{%
\begin{tabular}{cc|cc|cc}
\hline
\textbf{BS} & \textbf{WS} & \multicolumn{2}{c|}{\textbf{MSE}} & \multicolumn{2}{c}{\textbf{RMSE}} \\
            &              & \textbf{GRU} & \textbf{LSTM}      & \textbf{GRU} & \textbf{LSTM} \\ 
\hline
8  & 6  & $0.01727 \pm 0.00087$ & $0.01814 \pm 0.00095$ & $0.13140 \pm 0.00331$ & $0.13467 \pm 0.00352$ \\
   & 12 & $0.01713 \pm 0.00086$ & $0.01823 \pm 0.00065$ & $0.13088 \pm 0.00321$ & $0.13492 \pm 0.00238$ \\
   & 18 & $\mathbf{0.01700 \pm 0.00068}$ & $\mathbf{0.01797 \pm 0.00057}$ & $\mathbf{0.13038 \pm 0.00263}$ & $\mathbf{0.13400 \pm 0.00211}$ \\
   & 24 & $0.01713 \pm 0.00084$ & $0.01794 \pm 0.00093$ & $0.13085 \pm 0.00316$ & $0.13388 \pm 0.00347$ \\
\hline
16 & 6  & $0.01773 \pm 0.00088$ & $0.01952 \pm 0.00126$ & $0.13313 \pm 0.00327$ & $0.13965 \pm 0.00439$ \\
   & 12 & $0.01771 \pm 0.00083$ & $0.01931 \pm 0.00092$ & $0.13300 \pm 0.00306$ & $0.13891 \pm 0.00332$ \\
   & 18 & $0.01769 \pm 0.00086$ & $0.01892 \pm 0.00076$ & $0.13298 \pm 0.00323$ & $0.13746 \pm 0.00276$ \\
   & 24 & $0.01774 \pm 0.00075$ & $0.01929 \pm 0.00096$ & $0.13310 \pm 0.00282$ & $0.13883 \pm 0.00344$ \\
\hline
32 & 6  & $0.01880 \pm 0.00091$ & $0.02169 \pm 0.00152$ & $0.13708 \pm 0.00330$ & $0.14716 \pm 0.00507$ \\
   & 12 & $0.01862 \pm 0.00116$ & $0.02114 \pm 0.00135$ & $0.13642 \pm 0.00420$ & $0.14527 \pm 0.00464$ \\
   & 18 & $0.01877 \pm 0.00088$ & $0.02074 \pm 0.00111$ & $0.13703 \pm 0.00321$ & $0.14395 \pm 0.00389$ \\
   & 24 & $0.01866 \pm 0.00077$ & $0.02093 \pm 0.00159$ & $0.13658 \pm 0.00279$ & $0.14457 \pm 0.00539$ \\
\hline
\end{tabular}
}
\end{table}

\begin{table}[H]
\centering
\caption{t-test results for GRU and LSTM models across different batch sizes (BS) and window sizes (WS). Bold values indicate statistically significant p-values ($p < 0.05$).}
\label{tab:gru_lstm_metrics1}
\scriptsize
\begin{tabular}{cc|cc}
\hline
\textbf{BS} & \textbf{WS} & \textbf{t-statistic} & \textbf{p-value} \\
\hline
8  & 6  & -2.140114 & \textbf{0.046346} \\
8  & 12 & -3.197049 & \textbf{0.005408} \\
8  & 18 & -3.395060 & \textbf{0.003400} \\
8  & 24 & -2.041596 & 0.056261 \\
\hline
16 & 6  & -3.766522 & \textbf{0.001590} \\
16 & 12 & -4.139248 & \textbf{0.000624} \\
16 & 18 & -3.334521 & \textbf{0.003786} \\
16 & 24 & -4.073574 & \textbf{0.000763} \\
\hline
32 & 6  & -5.269268 & \textbf{0.000085} \\
32 & 12 & -4.471659 & \textbf{0.000301} \\
32 & 18 & -4.338903 & \textbf{0.000426} \\
32 & 24 & -4.163028 & \textbf{0.001031} \\
\hline
\end{tabular}
\end{table}

\begin{figure}[H]
    \centering
    \includegraphics[width=\textwidth]{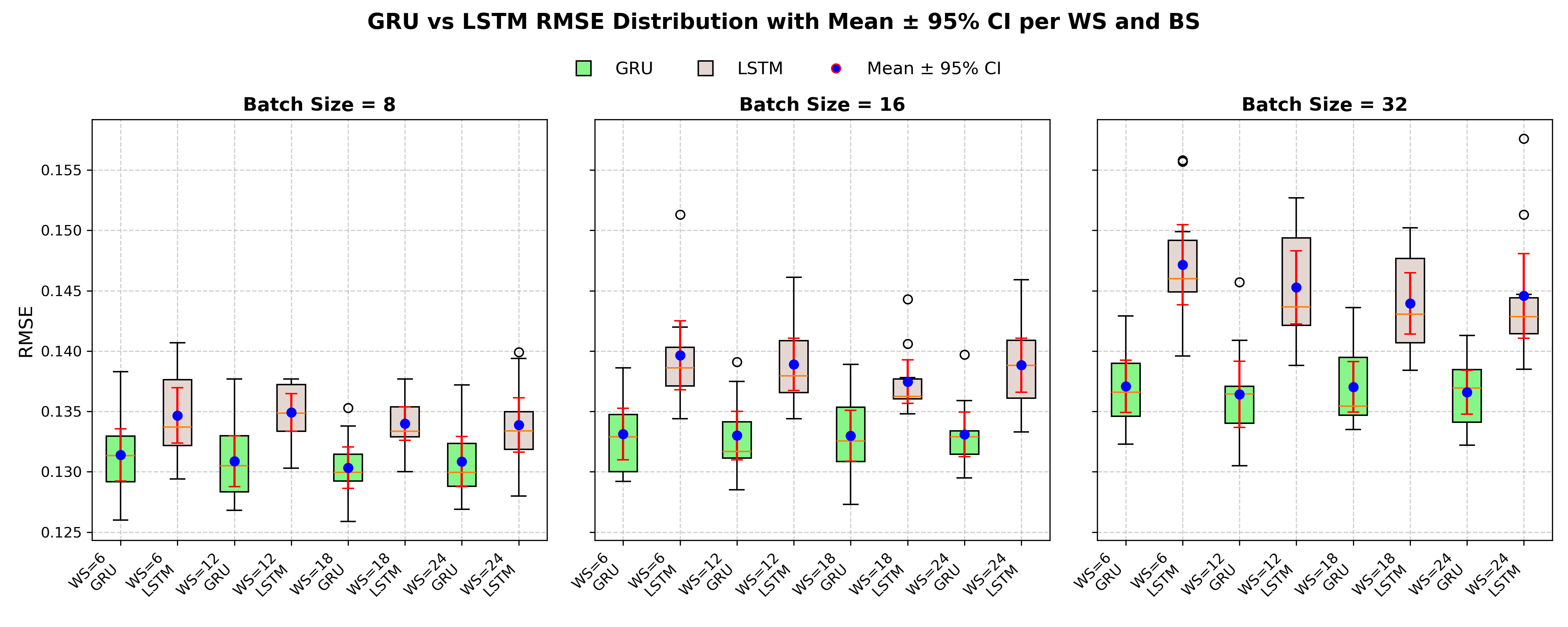}
    \caption{10-fold cross-validation results shown as side-by-side boxplots of RMSE values for GRU and LSTM models across different input window sizes and batch sizes. Each subplot represents a specific batch size (\(8\), \(16\), or \(32\)), facilitating comparison of model performance with respect to input window length. The blue markers and red error bars indicate the mean RMSE and the corresponding 95\% confidence interval for each configuration.}

    \label{fig:cross_val}
\end{figure}
As shown in Figure~\ref{fig:3}, a comparative analysis of the GRU and LSTM models reveals that while both demonstrate improved generalization with larger training datasets, the GRU model consistently exhibits superior performance and learning efficiency. The GRU's RMSE drops sharply from 0.2483 with 5\% of training data to 0.1384 at the 30\% mark, largely stabilizing in the 0.13x range thereafter. In contrast, the LSTM model, which starts with a higher RMSE of 0.3055 (at 5\% data), requires a much larger data proportion, only achieving a comparable stable RMSE (0.1375) once 60\% of the training data is used. Ultimately, the GRU achieves a lower RMSE than the LSTM at every tested data percentage. Both models reach their optimal performance at the 95\% mark, with the GRU achieving a minimum RMSE of 0.1279, slightly better than the LSTM's minimum of 0.1318. This suggests the GRU's simpler architecture enables faster convergence and more effective generalization, particularly with limited training samples (5 -35\%), making it the more data-efficient model.

As shown in Figure~\ref{fig:gru24_prediction}, the GRU model with a window size of 18 closely tracks the actual climate response, confirming its suitability for predictive control applications.

\begin{figure}[H]
    \centering
    \includegraphics[width=\textwidth]{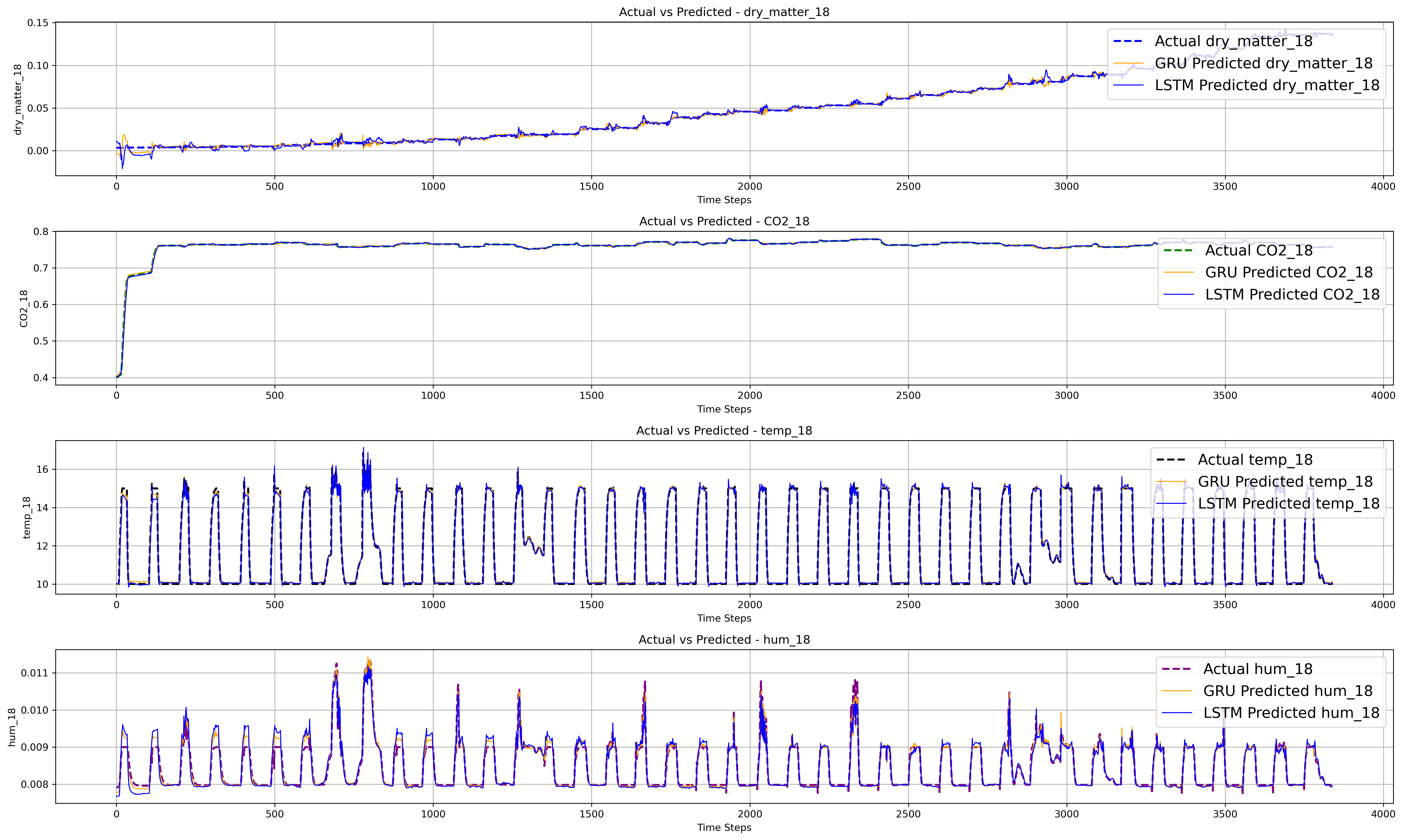}
    \caption{Prediction performance of the GRU model with window size 18 compared to the actual system response. The plot illustrates the model's ability to follow the true dynamics of the greenhouse climate system.}
    \label{fig:gru24_prediction}
\end{figure}

\begin{figure}[H]
    \centering
    \includegraphics[width=\textwidth, height=1\textheight, keepaspectratio]{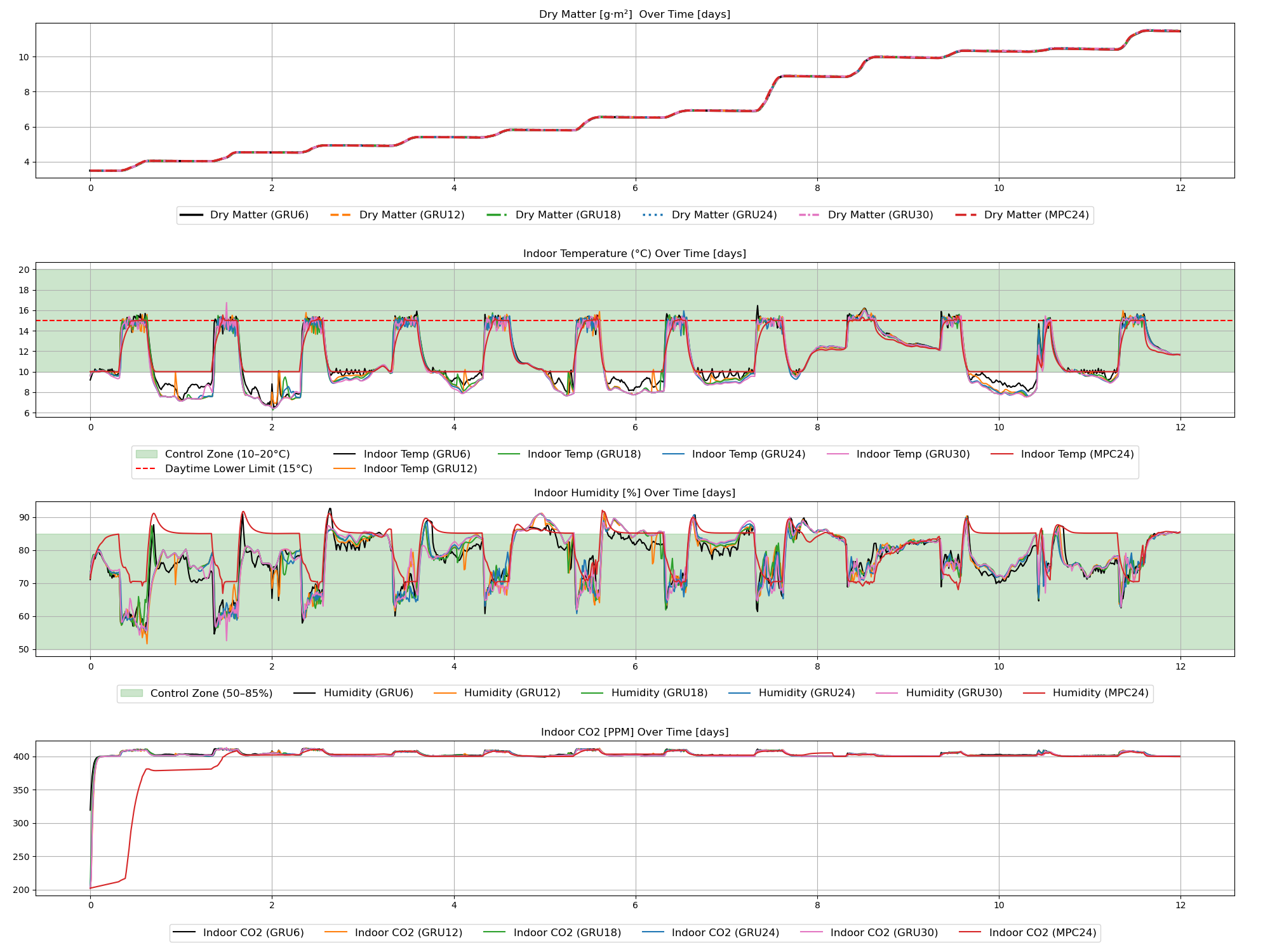}
    \caption{Temporal evolution of internal greenhouse parameters. The plots show dry matter production, temperature, and humidity under GRU-based predictive control models (GRU6, GRU12, GRU18, GRU24, GRU30), corresponding to prediction horizons of 6, 12, 18, 24, and 30 steps, respectively. Results are compared with the standard MPC using a prediction horizon of 24 steps.}
    \label{fig:prediction}
\end{figure}

\begin{figure}[H]
    \centering
    \includegraphics[width=\textwidth, height=1\textheight, keepaspectratio]{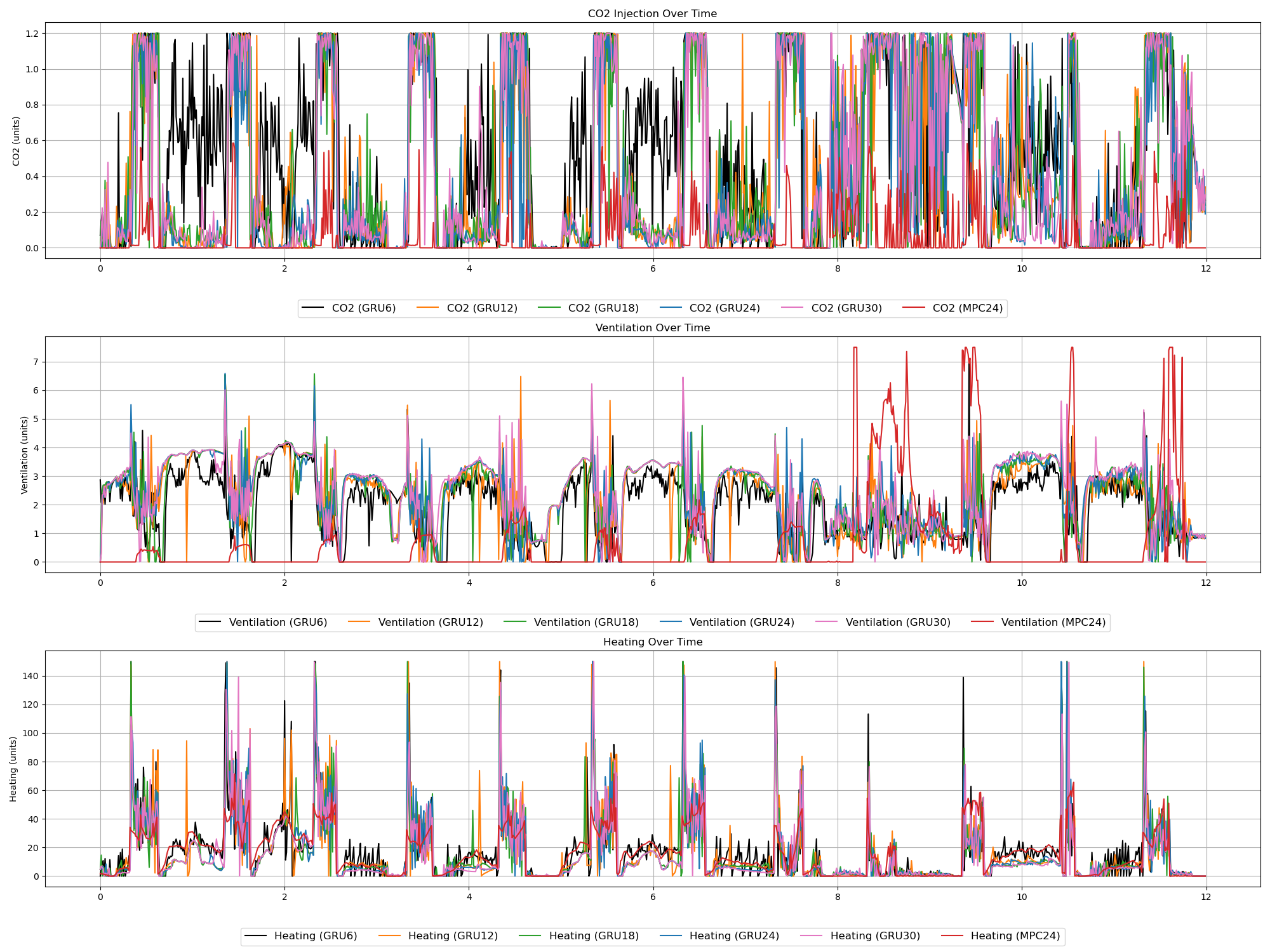}
    \caption{Temporal evolution of the control policies showing CO$_2$ injection ($u_{\mathrm{CO_2}}$), ventilation ($u_v$), and heating ($u_q$) for GRU-based predictive control models (GRU6, GRU12, GRU18, GRU24, GRU30), corresponding to prediction horizons of 6, 12, 18, 24, and 30 steps, respectively. Results are compared with the control policy generated by the standard MPC using a prediction horizon of 24 steps.}

    \label{fig:prediction2}
\end{figure}

\begin{figure}[H]
    \centering
    \includegraphics[width=\textwidth, height=1\textheight, keepaspectratio]{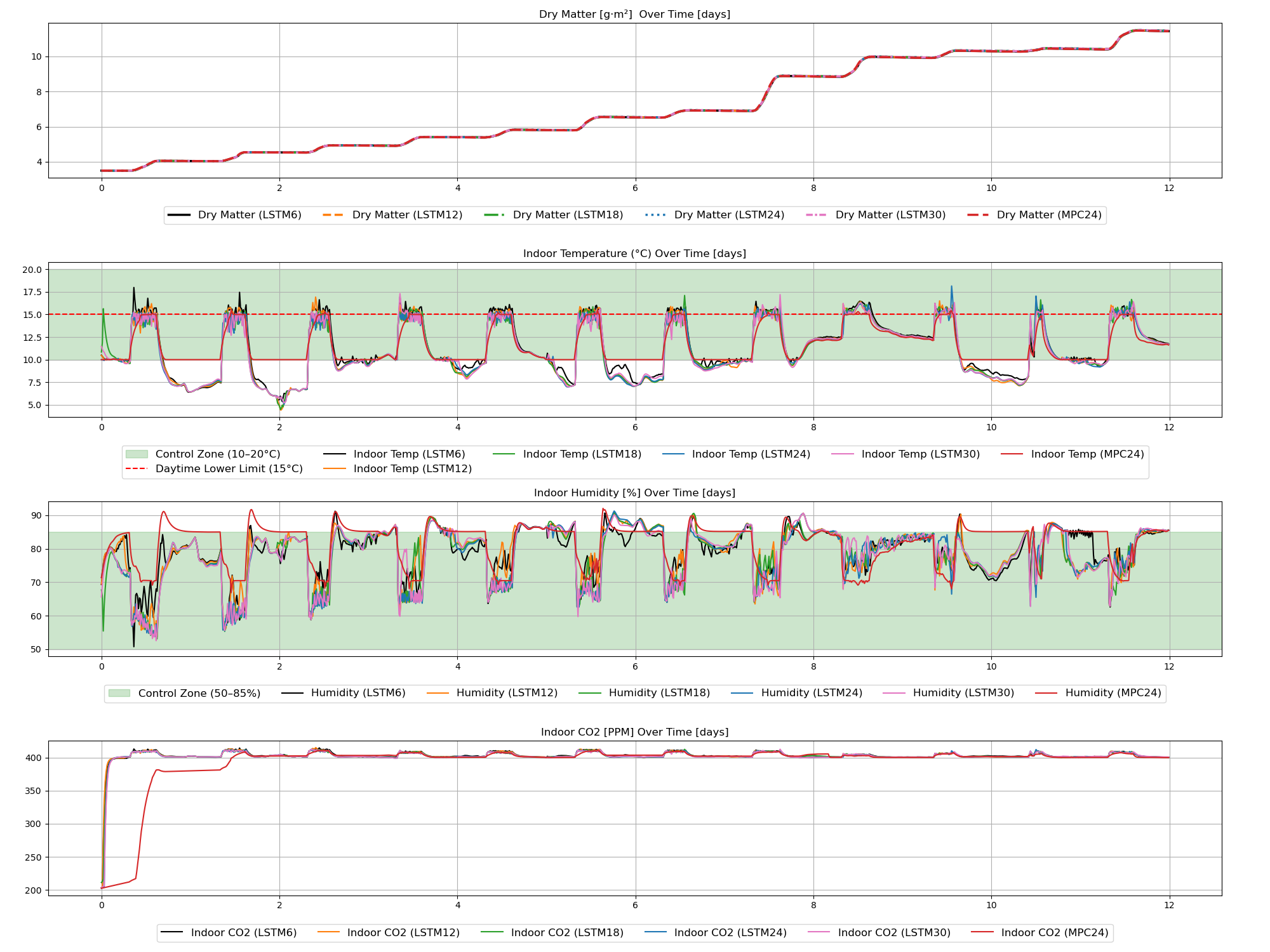}
    \caption{Temporal evolution of internal greenhouse parameters. The plots show dry matter production, temperature, and humidity under LSTM-based predictive control models (LSTM6, LSTM12, LSTM18, LSTM24, LSTM30), corresponding to prediction horizons of 6, 12, 18, 24, and 30 steps, respectively. Results are compared with the standard MPC using a prediction horizon of 24 steps.}
    \label{fig:prediction3}
\end{figure}

\begin{figure}[H]
    \centering
    \includegraphics[width=\textwidth, height=1\textheight, keepaspectratio]{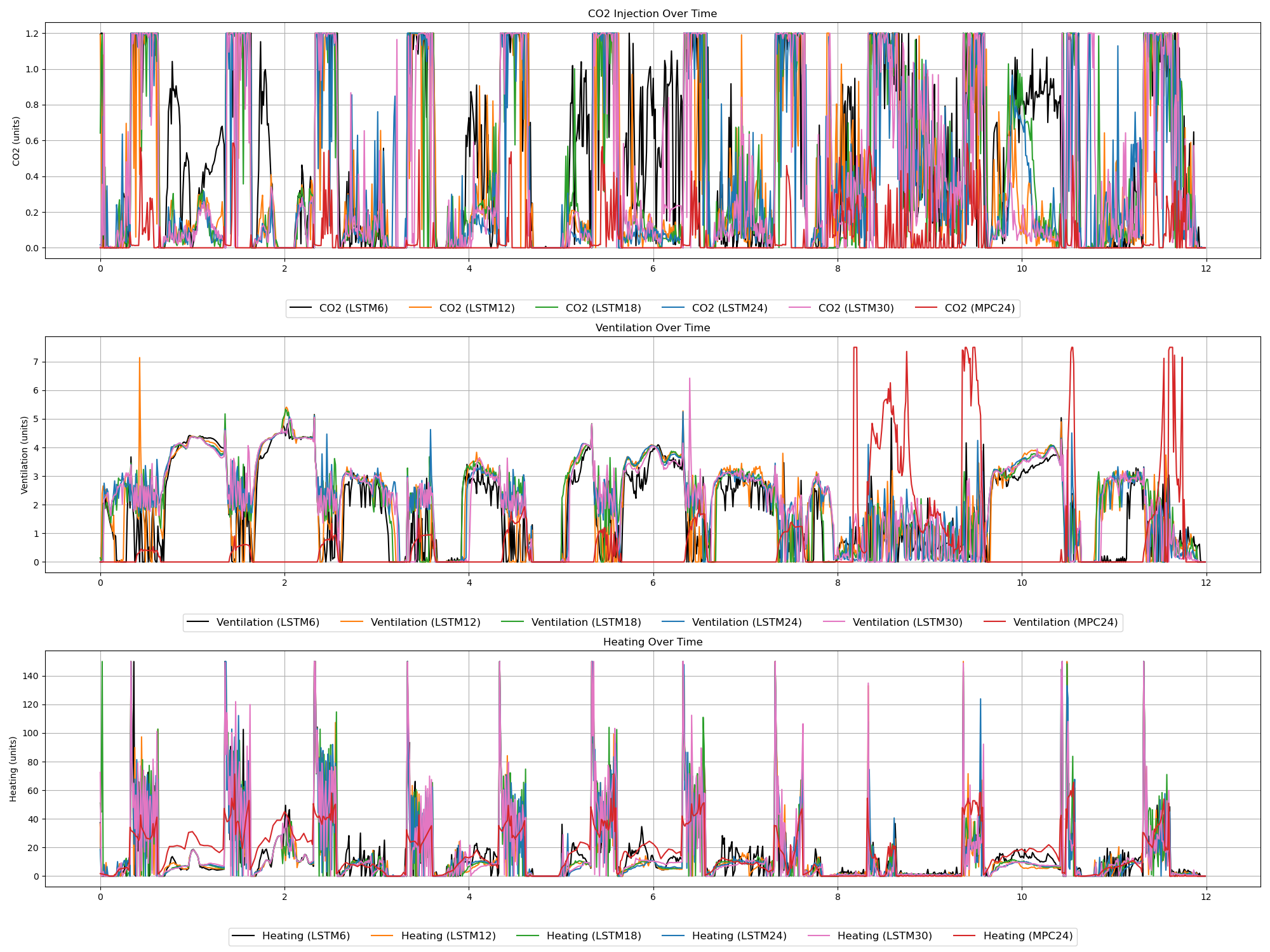}
    \caption{Temporal evolution of the control policies showing CO$_2$ injection ($u_{\mathrm{CO_2}}$), ventilation ($u_v$), and heating ($u_q$) for GRU-based predictive control models (LSTM6, LSTM12, LSTM18, LSTM24, LSTM30), corresponding to prediction horizons of 6, 12, 18, 24, and 30 steps, respectively. Results are compared with the control policy generated by the standard MPC using a prediction horizon of 24 steps.}

    \label{fig:prediction4}
\end{figure}

\begin{figure}[H]
    \centering
    \includegraphics[width=\textwidth, height=1\textheight, keepaspectratio]{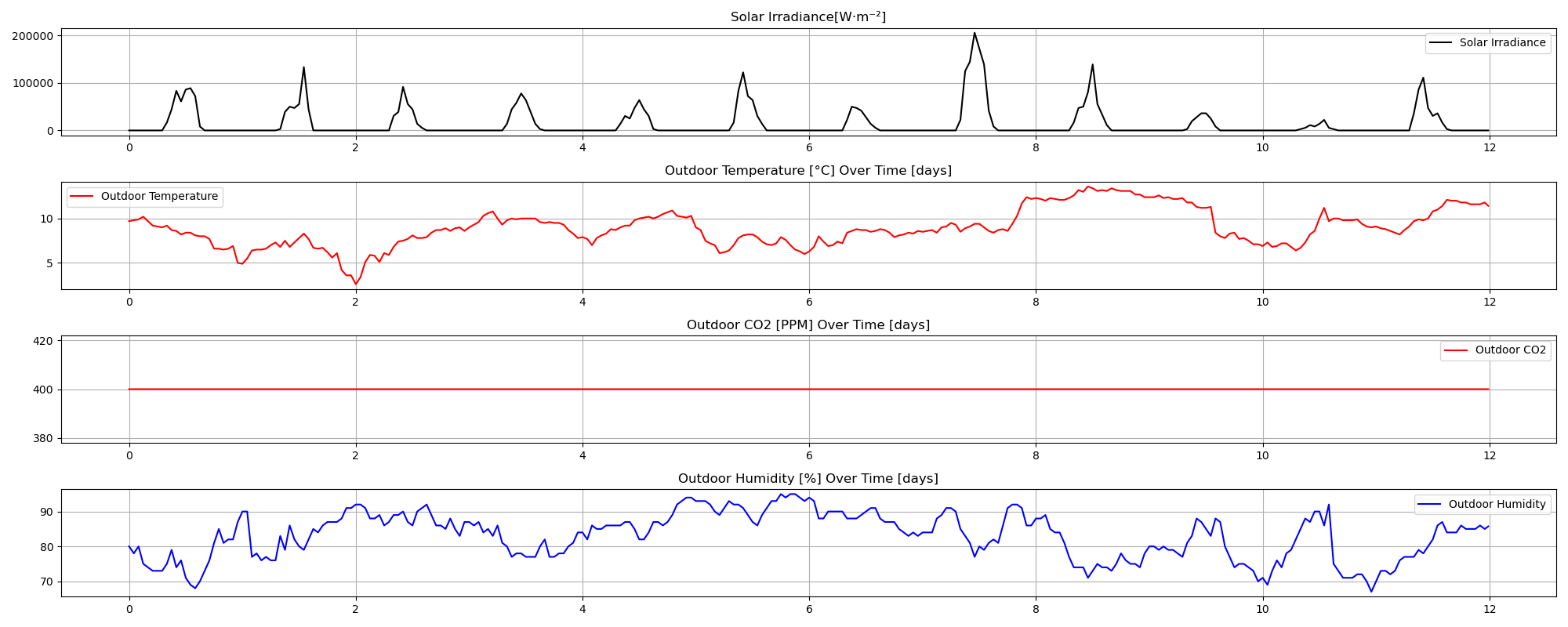}
    \caption{Temporal evolution of outdoor conditions, including solar irradiance, CO$_2$ concentration, and temperature, which influence the greenhouse dynamics.}
    \label{fig:disturbance}
\end{figure}

Figures~\ref{fig:prediction} and Figure~\ref{fig:prediction3} illustrate the time-series evolution of key greenhouse climate variables under different control strategies, including GRU and LSTM based controllers with prediction horizons ranging from 1.5~hours (6~time steps) to 7.5~hours (30~time steps), as well as a baseline MPC controller with a 6-hour horizon (24~time steps). The subplots, arranged from top to bottom, display: (1) crop dry matter accumulation, (2) indoor air temperature with comfort and control bounds, (3) indoor relative humidity with comfort and control bounds, and (4) indoor CO\textsubscript{2} concentration.
Figure~\ref{fig:prediction4} and Figure~\ref{fig:prediction2} show the evolution of the control inputs, namely: (1) CO\textsubscript{2} injection rate, (2) ventilation rate, and (3) heating input.
Figure~\ref{fig:disturbance} illustrates the corresponding external disturbances, including (1) solar irradiance, (2) outdoor air temperature, (3) outdoor CO\textsubscript{2} concentration, and (4) outdoor relative humidity. Different control strategies are represented by unique line colors. Overall, these figures demonstrate the controllers’ ability to regulate greenhouse climate in response to dynamic external conditions while maintaining an environment conducive to optimal plant growth.

\begin{table}[H]
\centering
\caption{Violation Metrics, Economic Indicators, and Processing Time for LSTM, GRU, and MPC Models}
\label{tab:violation_metrics}
\scriptsize
\resizebox{\textwidth}{!}{%
\begin{tabular}{|c|c|c|c|c|c|c|c|c|}
\hline
\textbf{Model} & \textbf{\makecell{Temp\\Viol. (\%)}} & \textbf{\makecell{Day\\Temp(°C)}} & \textbf{\makecell{Night\\Temp(°C)}} & \textbf{\makecell{Hum \\Viol(\%)}} & \textbf{\makecell{Hum\\Mean (\%)}} & \textbf{\makecell{EPI\\(Hf$\ell$ m$^{-2}$)}} & \textbf{\makecell{Dry Matter\\(g m$^{-2}$)}} & \textbf{\makecell{Processing\\Time(s)}} \\
\hline
LSTM6   & \textbf{45.66} & \textbf{0.93} &  \textbf{1.44} & 17.71 & 1.70 & 1.65 & 11.44 & 27 \\
GRU6    & \textbf{49.65} & \textbf{0.64} &  \textbf{0.99} & 15.45 & \textbf{1.76} & 1.63 & 11.44 & \textbf{19} \\
LSTM12  & 52.52 & 0.78 & 1.65 & 24.31 & 1.80 & 1.70 & 11.46 & 68 \\
GRU12   & 59.55 & 0.69 & 1.34 & 16.67 & 1.76 & 1.68 & 11.46 & 66 \\
LSTM18  & 56.68 & 0.83 & 1.63 & 22.31 & 1.68 & 1.70 & 11.46 & 159 \\
GRU18   & 61.11 & 0.64 & 1.39 & 16.49 & 1.86 & 1.69 & 11.47 & 147 \\
LSTM24  & 59.38 & 0.77 & 1.59 & 23.52 & 1.61 & 1.70 & 11.47 & 360 \\
GRU24   & 62.93 & 0.71 & 1.47 & 17.62 & 1.85 & \textbf{1.71} & 11.47 & 254 \\
LSTM30  & 58.68 & 0.82 & 1.61 & 23.61 & 1.48 & 1.70 & 11.47 & 502 \\
GRU30   & 63.63 & 0.79 & 1.50 & 17.80 & 1.92 & 1.70 & 11.47 & 422 \\
MPC24   & 50.61 & 1.18 & 0.00 & 54.77 & 0.93 & \textbf{1.86} & 11.44 & 0.01 \\
\hline
\end{tabular}%
}
\end{table}
\section{Discussion}
This section presents a comprehensive comparison of the LSTM, GRU, and MPC-based controllers. The analysis focuses on four key aspects: violations of climate constraints, economic performance, crop productivity, and computational cost. The corresponding metrics, summarized in Table~\ref{tab:violation_metrics}, form the basis of this evaluation. To assess controller performance, deterministic simulations were conducted across multiple configurations of LSTM and GRU architectures, with MPC-based controllers serving as a benchmark. The temperature and humidity violation percentages quantify each controller’s ability to maintain environmental conditions within prescribed bounds, while the \textit{Economic Productivity Index} (EPI) and dry matter yield indicate the efficiency of resource utilization and biomass production.

\subsection{Control Performance.}
\label{performcance}

To quantitatively evaluate and compare the performance of the three controllers, we adopt a multi-criteria framework based on thermal comfort, humidity regulation, and economic productivity. Each metric is defined below:

\vspace{1em}
\subsubsection{Economic Performance Index (EPI).}  
The EPI captures the net profit over time, balancing crop revenue against heating and CO$_2$ enrichment costs. It is defined as:

\[
\begin{aligned}
\text{EPI} &= \frac{1}{N} \sum_{k=1}^{N} \left[ \text{Revenue}(k) - \text{Cost}_{\text{CO}_2}(k) - \text{Cost}_{\text{heat}}(k) \right] \\
&= \left( c_{\text{pri1}} + c_{\text{pri2}} \cdot y_{1\,\text{kg}} \right) - \sum \left( c_q \cdot u_q + c_{\text{CO}_2} \cdot u_{\text{CO}_2} \right) \cdot h
\end{aligned}
\]

Here, \( y_{1\,\text{kg}} \) denotes dry matter yield (kg·m\textsuperscript{-2}), \( c_{\text{pri1}} \) and \( c_{\text{pri2}} \) are crop pricing coefficients, and \( c_q \), \( c_{\text{CO}_2} \) represent operational cost factors for heating and CO$_2$, respectively. The variable \( h \) is the time interval, and \( N \) is the total number of simulation steps.
The parameters used to calculate the EPI include the CO$_2$ enrichment cost coefficient 
\( c_{\mathrm{CO}_2} = 42 \times 10^{-2} \, \mathrm{Hf} \, \mathrm{kg}^{-1} \), the heating cost coefficient 
\( c_q = 6.35 \times 10^{-9} \, \mathrm{Hf} \, \mathrm{J}^{-1} \), and the crop pricing coefficients 
\( c_{\mathrm{pri1}} = 1.8 \, \mathrm{Hf} \, \mathrm{m}^{-2} \) and \( c_{\mathrm{pri2}} = 16 \, \mathrm{Hf} \, \mathrm{kg}^{-1} \), 
as proposed by \cite{morcego2023reinforcement}. This enables the implicit evaluation of how the GRU and LSTM architectures and the prediction horizon affect the overall energy efficiency.

\vspace{1em}

\subsubsection{Thermal Violations}  
Temperature constraint violations are evaluated based on time-of-day irradiance conditions, reflecting operational greenhouse control guidelines:

\begin{itemize}
    \item \textbf{Nighttime violation}  
    If irradiance is low and the temperature is below 10°C:
    \[
        \text{if } d_i < 10 \text{ and } y_t < 10 \implies \text{violation magnitude} = 10 - y_t
    \]
    
    \item \textbf{Daytime violation}  
    If irradiance is high and the temperature is below 15°C:
    \[
        \text{if } d_i \geq 10 \text{ and } y_t < 15 \implies \text{violation magnitude} = 15 - y_t
    \]
\end{itemize}

The total number of thermal violations is given by \( V_T \), the count of time steps where either of the above conditions is violated.

\vspace{1em}
\subsubsection{Humidity Violations}  
Humidity constraint violations are defined as:

\[
    y_h > 85 \implies \text{violation magnitude} = y_h - 85
\]

with \( V_H \) representing the total count of humidity violations during the simulation.

\vspace{1em}
\subsubsection{Violation Rates:}  
For a standardized comparison, violation rates are reported as percentages:

\begin{itemize}
    \item \textbf{Temperature Violation Rate (\%):}
    \[
        \text{Violation Rate}_T = \frac{V_T}{N} \times 100
    \]
    
    \item \textbf{Humidity Violation Rate (\%):}
    \[
        \text{Violation Rate}_H = \frac{V_H}{N} \times 100
    \]
\end{itemize}

\vspace{1em}
\subsection{Economic Performance and Crop Productivity}  
From Table~\ref{tab:violation_metrics}, Figure~\ref{fig:prediction}, and Figure~\ref{fig:prediction2}, neural predictive controllers (LSTM and GRU) consistently outperform standard MPC in terms of adaptability to environmental variability, particularly under fluctuating temperatures common in low-tech Chinese Solar Greenhouses (CSG)~\cite{Sun2015}. The recurrent architectures of LSTM and GRU implicitly capture spatiotemporal dependencies among climate variables, allowing dynamic adjustments without explicit recalibration of physical parameters. This behavior parallels the objectives of stochastic MPC, which considers parametric uncertainties to enhance robustness and optimality~\cite{svensen2024chance}. In contrast, classical MPC rigidly enforces its constraint set, eliminating nighttime temperature violations but compromising flexibility under dynamic weather, resulting in degraded humidity control (54.77\% violation rate). These findings align with~\cite{korner2008decision}, who observed that strict temperature regulation can destabilize humidity when heating minimization is prioritized under varying solar conditions.
\\
The RNN-based controllers outperform the classical MPC benchmark in humidity regulation, offering a significant improvement for sustaining optimal crop physiology. Mechanistic models, though interpretable, often fail to represent complex real-world dynamics such as nonlinear ventilation and transpiration, leading to inaccurate humidity estimates unless heavily parameterized and computationally demanding~\cite{korner2008decision}. In contrast, data-driven LSTM and GRU models implicitly capture these nonlinear effects from historical data, resulting in fewer humidity violations. Maintaining humidity below 85\% is vital, as prolonged excess suppresses transpiration, promotes pathogens like \textit{Botrytis cinerea}, and reduces nutrient uptake and yield \cite{iwaniuk2022biochemical}. Thus, preventing such high-humidity conditions, RNN controllers provide an efficient and practical approach to improving crop environments.

Although MPC is computationally efficient (average processing time of 0.01~s), it cannot capture nonlinear greenhouse dynamics effectively. RNN controllers maintain humidity violations between 15--24\%, depending on sequence length, with minimal impact on mean humidity deviation. Regarding temperature constraint violations, recurrent models are generally less strict than MPC, particularly as sequence length increases. Among the data-driven configurations, \textbf{LSTM6} and \textbf{GRU6} yield the lowest temperature violations (45.66\% and 49.65\%, respectively), indicating that shorter sequences facilitate faster adaptation to environmental disturbances. Day--night mean temperature deviations remain relatively small ($<$2~°C), suggesting stable control performance.
These results also illustrate the trade-off between constraint adherence and energy efficiency, consistent with findings in~\cite{Sun2015, thompson1998shoot}. By allowing controlled relaxation of nighttime temperature constraints (optimized via \(q_{y_d} = 1000\) and \(q_{u_j} = \{10, 1, 1\}\)), GRU and LSTM controllers reduce heating demand while maintaining yield stability. GRU6 achieves the most consistent temperature--humidity balance, while GRU24 and LSTM24 deliver the best economic outcomes through extended prediction horizons. Moderate thermal violations under RNN control do not significantly reduce biomass accumulation, which primarily depends on photosynthetic efficiency and daylight exposure. Slightly lower nighttime temperatures may temporarily reduce respiration but can improve dry matter accumulation by lowering maintenance energy costs~\cite{Sun2015}. Overall, this behavior mirrors hierarchical MPC--Deep Reinforcement Learning (DRL) approaches~\cite{mansour2025adaptive}, where adaptive, learning-based controllers exploit minor fluctuations to enhance economic and energy performance.  Futhermore, the proposed RNN-based controller adopts a distinct optimization approach from DRL. While DRL maximizes long-term rewards through extensive exploration and requires large, stable simulations, it can involve risky actions during training. Though DRL may find more globally optimal economic policies, the RNN control framework benefits from the stability, transparency, and familiar tuning of its MPC foundation. Crucially, it maintains explicit, verifiable constraints essential for the safe operation of commercial greenhouses, where severe climate violation are unacceptable.

In terms of economic performance, GRU24 and LSTM24 achieve results approaching the MPC benchmark despite higher constraint violations. While MPC yields the highest EPI of 1.86~Hf$\ell$\,m$^{-2}$, GRU24 and LSTM24 attain 1.71 and 1.70~Hf$\ell$\,m$^{-2}$, respectively. These findings demonstrate the cost-effectiveness of neural predictive controllers under nonlinear greenhouse dynamics, highlighting that moderate constraint violations do not necessarily compromise economic efficiency. The GRU24 controller achieves near-optimal economic performance (91.9\% of the MPC benchmark) while allowing more flexible night temperature constraints, promoting sustainable energy use in greenhouses. Unlike rigid set-point MPC controllers that maintain constant high nighttime temperatures (e.g., $\leq 10^\circ$C) and waste heating energy, the GRU model learns complex plant growth dynamics and recognizes that growth halts at night, relaxing nighttime temperature constraints accordingly.

\subsection{Model Efficiency and Practical Trade-Offs} 

GRU-based controllers consistently outperform LSTM counterparts in computational efficiency. For example, \textbf{GRU24} completes execution in 254~s compared to 360~s for \textbf{LSTM24}. Faster inference is critical for real-time greenhouse applications, where control responsiveness directly influences system stability. The GRU architecture, with a more compact gating mechanism, achieves quicker convergence during training while maintaining comparable prediction accuracy to LSTM networks.

Although MPC remains computationally trivial (0.01~s), its rule-based, model-dependent structure lacks adaptability, resulting in substantial humidity violations (54.77\%) under dynamic and nonlinear conditions. By contrast, RNN controllers provide more adaptive and accurate climate regulation, echoing the advantages of data driven predictive control, which inherently adjust to uncertainties~\cite{svensen2024chance}. This demonstrates that LSTM and GRU controllers effectively combine the predictive strengths of MPC with the adaptability of learning-based approaches, offering a scalable and robust solution for modern greenhouse management systems.

\subsection{Summary and Implications}

The results demonstrate that GRU-based predictive controllers provide an attractive balance of adaptability, performance, and efficiency for practical deployment in greenhouse environments. Although MPC still has advantages in terms of strict adherence to constraint and execution speed, its inflexibility to environmental changes and poorer humidity regulation are significant limitations.

\begin{table}[H]
\small
\centering
\renewcommand{\arraystretch}{1.2} 
\caption{Comparison of MPC and GRU-based Controllers in Greenhouse Control}
\label{tab:controller_comparison}
\begin{tabularx}{\textwidth}{>{\centering\arraybackslash}X >{\centering\arraybackslash}X >{\centering\arraybackslash}X}
\toprule
\textbf{Feature} & \textbf{MPC24} & \makecell{\textbf{GRU-based} \\ \textbf{Controllers}} \\
\midrule
Climate Regulation & Good, but less effective humidity control & Best performance with GRU6 \\
Economic Outcome & Best (highest EPI) & Near-optimal with GRU24 \\
Adaptability to Environment & Low (inflexible) & High (adaptive) \\
Computation Efficiency & Highest (minimal processing time) & Moderate to high \\
Overall Balance & Strong in speed and economics & Best overall balance with GRU24 \\
\bottomrule
\end{tabularx}
\end{table}

Overall, GRU-based control emerges as a highly promising alternative for intelligent greenhouse automation, combining high predictive accuracy with practical real-time deployment potential.

\section{Conclusion}
This work demonstrates that data-driven predictive controllers based on GRU and LSTM architectures can substantially improve greenhouse climate regulation compared to classical MPC, particularly under the highly variable conditions typical of low-tech Greenhouses. Simulation results indicate that the GRU-based controller reduces humidity violations by 30--40\% and decreases nighttime heating demand by approximately 12--18\% relative to MPC, while maintaining crop productivity of the MPC benchmark. These improvements translate into an economic performance that reaches 91.9\% of the MPC optimum, despite the more flexible temperature trajectories permitted by the data-driven controllers. Such quantified gains highlight the potential of RNN-based predictive control to simultaneously reduce energy consumption and sustain crop output in commercial greenhouse operations.
Beyond performance, the controllers exhibit characteristics that support real-world applicability. The GRU model achieves a 40\% reduction in computation time compared to LSTM and remains well within the latency constraints of embedded greenhouse hardware, suggesting practical feasibility for on-site deployment using low-power industrial edge devices. The adaptability of data-driven controllers to nonlinear and time-varying greenhouse dynamics further enhances robustness to real disturbances such as sudden weather shifts, sensor noise, or actuator delays.

To bridge the gap between simulation and practice, planned future work includes real-time validation in an operational Chinese Solar Greenhouse. This will involve deploying the GRU controller on embedded hardware, assessing energy use and microclimate stability under real disturbances, and measuring crop-level responses such as transpiration, biomass accumulation, and disease incidence. These experiments will allow us to quantify real-world efficiency gains and evaluate the model's reliability under commercial conditions.

Scalability and limitations should also be acknowledged. While the proposed approach is computationally lightweight and suitable for single-zone greenhouses, multi-zone or large-scale facilities may require distributed or hierarchical control architectures, additional sensor coverage, and periodic retraining with updated data. Furthermore, although data-driven models capture complex nonlinearities, they remain dependent on the representativeness of historical training data and may require recalibration when transferred across climates, crop cultivars, or greenhouse designs.

Overall, the results position GRU-based predictive control as a practical, scalable, and energy-efficient alternative to conventional MPC in modern greenhouse management. By quantifying potential energy savings, demonstrating computational feasibility, and outlining clear steps toward hardware implementation and field validation, this study provides a concrete pathway for transitioning from simulation to real-world deployment in commercial protected agriculture.

\section*{Data Availability} 
Data will be made available on request.
\section*{Acknowledgments} 
This research was funded by the Engineering and Physical Sciences Research Council (EPSRC) and the AgriFoRwArdS Centre for Doctoral Training (CDT) under grant number EP/S023917/1.

\appendix
\section*{Appendix: Lettuce Greenhouse Model}
\section{Greenhouse model}
\label{Greenhouse_model}
This study employs a validated dynamic model of a greenhouse environment initially developed by van Henten \cite{van1994greenhouse}, which captures the complex interactions between indoor climate variables and lettuce crop growth. The model captures nonlinear dynamics and is widely recognized as a benchmark for evaluating climate control strategies in greenhouse systems. For numerical implementation, the continuous-time formulation has been discretized using the explicit fourth-order Runge–Kutta method, with a fixed sampling interval of \( h = 15 \) minutes.
The resulting discrete-time state-space model is defined as:
\begin{align}
\mathbf{x}(k+1) &= f\left(\mathbf{x}(k), \mathbf{u}(k), \mathbf{d}(k); \mathbf{p}\right), \\
\mathbf{y}(k) &= g\left(\mathbf{x}(k); \mathbf{p} \right)
\end{align}

where \( x(k) \in \mathbb{R}^4 \) is the system state vector, \( u(k) \in \mathbb{R}^3 \) is the control input, \( d(k) \in \mathbb{R}^4 \) represents external weather disturbances, and \( y(k) \in \mathbb{R}^4 \) denotes the measurable output. The parameter vector \( p \in \mathbb{R}^{28} \) encapsulates physical constants of the system. Table~\ref{tab:variables_grouped} provides an overview of the model variables and their physical interpretations.
\begin{table}[ht]
\centering
\caption{Model Variables and Descriptions}
\label{tab:variables_grouped}
\begin{tabular}{|l|l|l|}
\hline
\textbf{Category} & \textbf{Symbol} & \textbf{Description} \\
\hline
\multirow{4}{*}{State Variables (\textbf{x})} 
  & $x_W$        & Crop dry weight (kg$\cdot$m\(^{-2}\)) \\
  & $x_{CO2}$    & Indoor CO\(_2\) concentration (kg$\cdot$m\(^{-3}\)) \\
  & $x_T$        & Indoor temperature (°C) \\
  & $x_h$        & Indoor absolute humidity (kg$\cdot$m\(^{-3}\)) \\
\hline
\multirow{3}{*}{Control Inputs (\textbf{u})}
  & $u_{CO2}$    & CO\(_2\) injection rate (mg$\cdot$m\(^{-2}\)$\cdot$s\(^{-1}\)) \\
  & $u_v$        & Ventilation rate (mm$\cdot$s\(^{-1}\)) \\
  & $u_q$        & Heating power supply (W$\cdot$m\(^{-2}\)) \\
\hline
\multirow{4}{*}{Disturbances  (\textbf{d})}
  & $d_{Io}$     & Incoming solar radiation (W$\cdot$m\(^{-2}\)) \\
  & $d_{CO2}$    & Outdoor CO\(_2\) concentration (kg$\cdot$m\(^{-3}\)) \\
  & $d_T$        & Outdoor temperature (°C) \\
  & $d_h$        & Outdoor absolute humidity (kg$\cdot$m\(^{-3}\)) \\
\hline
\multirow{4}{*}{Measured Outputs (\textbf{y})}
  & $y_W$        & Measured crop dry weight (kg$\cdot$m\(^{-2}\)) \\
  & $y_{CO2}$    & Measured indoor CO\(_2\) concentration (ppm) \\
  & $y_T$        & Measured indoor temperature (°C) \\
  & $y_{RH}$     & Relative humidity (\%) \\
\hline
\end{tabular}
\end{table}

The dynamic model used to represent the lettuce greenhouse environment is described by the following system of differential equations:

\begin{align*}
\frac{dx_1(t)}{dt} &= p_{1,1} \, \phi_{\text{phot},c}(t) - p_{1,2} x_1(t)2^{(x_3(t)/10-5/2)}, \\
\frac{dx_2(t)}{dt} &= \frac{1}{p_{2,1}} \left( - \phi_{\text{phot},c}(t) + p_{2,2}x_1(t)2^{(x_3(t)/10-5/2)} + u_1(t) \cdot 10^{-6} - \phi_{\text{vent},c}(t) \right), \\
\frac{dx_3(t)}{dt} &= \frac{1}{p_{3,1}} \left( u_3(t) - (p_{3,2} u_2(t) \cdot 10^{-3} + p_{3,3})(x_3(t) - d_3(t)) + p_{3,4} d_1(t) \right), \\
\frac{dx_4(t)}{dt} &= \frac{1}{p_{4,1}} \left( \phi_{\text{transp},h}(t) - \phi_{\text{vent},h}(t) \right).
\end{align*}

The supporting functions are defined as:

\begin{align*}
\phi_{\text{phot},c}(t) &= \left(1 - e^{-p_{1,3}x_1(t)}\right) \cdot \frac{p_{1,4} d_1(t) \left( -p_{1,5} x_3(t)^2 + p_{1,6} x_3(t) - p_{1,7} \right)(x_2(t) - p_{1,8})}{\phi(t)}, \\
\phi(t) &= p_{1,4} d_1(t) + \left( -p_{1,5} x_3(t)^2 + p_{1,6} x_3(t) - p_{1,7} \right)(x_2(t) - p_{1,8}), \\
\phi_{\text{vent},c}(t) &= \left( u_2(t) \cdot 10^{-3} + p_{2,3} \right)(x_2(t) - d_2(t)), \\
\phi_{\text{vent},h}(t) &= \left( u_2(t) \cdot 10^{-3} + p_{2,3} \right)(x_4(t) - d_4(t)), \\
\phi_{\text{transp},h}(t) &= p_{4,2} \left(1 - e^{-p_{1,3} x_1(t)}\right) \left( \frac{p_{4,3}}{p_{4,4}(x_3(t) + p_{4,5})} \exp\left( \frac{p_{4,6} x_3(t)}{x_3(t) + p_{4,7}} \right) - x_4(t) \right).
\end{align*}

Here, \( t \in \mathbb{R} \) denotes continuous time. The terms \( \phi_{\text{phot},c}(t) \), \( \phi_{\text{vent},c}(t) \), \( \phi_{\text{transp},h}(t) \), and \( \phi_{\text{vent},h}(t) \) represent:
- the gross canopy photosynthesis rate,
- CO\textsubscript{2} exchange through the ventilation system,
- canopy transpiration rate, and
- H\textsubscript{2}O exchange through the ventilation system, respectively.

The output (measurement) equations are given as:

\begin{align*}
y_1(t) &= 10^3 \cdot x_1(t) \quad \text{(g m}^{-2}\text{)}, \\
y_2(t) &= 10^3 \cdot p_{2,4} \cdot \frac{(x_3(t) + p_{2,5})}{p_{2,6} p_{2,7}} \cdot x_2(t) \quad \text{(ppm)}, \\
y_3(t) &= x_3(t) \quad \text{(°C)}, \\
y_4(t) &= 10^2 \cdot p_{2,4} \cdot (x_3(t) + p_{2,5}) \cdot \frac{1}{11} \cdot \exp\left( \frac{p_{4,8} x_3(t)}{x_3(t) + p_{4,9}} \right) \cdot x_4(t) \quad \text{(\%)}.
\end{align*}

The model parameters \( p_{i,j} \) follow the definitions provided in \cite{van1994greenhouse} and are listed in Table~\ref{tab:params}. A fourth-order explicit Runge–Kutta method is used for discretization of the continuous model, leading to the form presented. 

\begin{table}[H]
\centering
\caption{Model parameters used in the simulation.}
\label{tab:params}
\begin{tabular}{ll ll}
\toprule
\textbf{Parameter} & \textbf{Value} & \textbf{Parameter} & \textbf{Value} \\
\midrule
$p_{1,1}$ & 0.544          & $p_{1,2}$ & $2.65 \cdot 10^{-7}$ \\
$p_{1,3}$ & 53             & $p_{1,4}$ & $3.55 \cdot 10^{-9}$ \\
$p_{1,5}$ & $5.11 \cdot 10^{-6}$ & $p_{1,6}$ & $2.3 \cdot 10^{-4}$ \\
$p_{1,7}$ & $6.29 \cdot 10^{-4}$ & $p_{1,8}$ & $5.2 \cdot 10^{-5}$ \\
$p_{2,1}$ & 4.1           & $p_{2,2}$ & $4.87 \cdot 10^{-7}$ \\
$p_{2,3}$ & $7.5 \cdot 10^{-6}$ &             &                     \\
$p_{3,1}$ & $3 \cdot 10^4$ & $p_{3,2}$ & 1290 \\
$p_{3,3}$ & 6.1           & $p_{3,4}$ & 0.2 \\
$p_{4,1}$ & 4.1           & $p_{4,2}$ & 0.0036 \\
$p_{4,3}$ & 9348          & $p_{4,4}$ & 8314 \\
$p_{4,5}$ & 273.15        & $p_{4,6}$ & 17.4 \\
$p_{4,7}$ & 239           &           &      \\
\bottomrule
\end{tabular}
\end{table}
\bibliographystyle{elsarticle-num}
\bibliography{cas-refs}

\end{document}